\documentclass[prl,twocolumn,epsf,psfig]{revtex4}
\usepackage{graphicx}
\def\widetext{\end{twocolumn}}
\input epsf

\begin{document}

\title{Profiles of near-resonant population-imbalanced trapped Fermi gases}

\author{Theja N. De Silva$^{a,b}$ and Erich J. Mueller$^{a}$}
\affiliation{$^{a}$ LASSP, Cornell University, Ithaca, New York 14853, USA. \\
$^{b}$ Department of physics, University of Ruhuna, Matara, Sri
Lanka.}

\begin{abstract}
We investigate the density profiles of a partially polarized trapped
Fermi gas in the BCS-BEC crossover region using mean field theory
within the local density approximation. Within this approximation
the gas is phase separated into concentric shells.  We describe how
the structure of these shells depends upon the polarization and the
interaction strength. A comparison with experiments yields insight
into the possibility of a polarized
superfluid phase. \\

\end{abstract}

\maketitle

\section{Introduction}

Advances in the trapping and manipulating of degenerate Fermi atoms
are attracting intense interest from physicists in the fields of
condensed matter physics, atomic molecular and optical physics,
nuclear physics, astrophysics, and particle physics. Current and
future experiments aim to use this highly controlled environment to
explore many-body phenomena with impact on widely varying areas of
physics. We discuss the theory of one such set of phenomena; the
properties of trapped partially polarized fermionic atoms, where the
two spin states have different Fermi surfaces.

The study of Fermi systems with mismatched Fermi surfaces began with
attempts by Fulde and Ferrell and Larkin and Ovchinnikov
(FFLO)\cite{fflo} to understand magnetized superconductors.  More
recent work has come from researchers in the nuclear physics
community who are studying superconductivity in nuclear matter and
quark matter, with a possible application to neutron stars or heavy
ion collisions \cite{nm}.  Such calculations have taken on new
relevance with the possibility of cold gas experiments where alkali
atoms are trapped in a number of distinguishable hyperfine states,
with negligible spin relaxation. Thus one can produce a
two-component Fermi gas with arbitrary population imbalance. Using
magnetic-field driven Feshbach resonances \cite{feshbach}, the
interactions between these atoms can be made large enough to drive
the system superfluid.

Very recently, there have been two experimental studies of trapped
$^6$Li Fermi atoms with mismatched Fermi surfaces
\cite{ketterle,randy}. By analyzing time-of-flight images, Zwierlein
\emph{et al.} \cite{ketterle} found evidence for phase separation
between regions of equal and unequal population density.
Furthermore, by studying vortices, they were able to monitor the
evolution of superfluidity as a function of population imbalance:
finding not only that polarization inhibits superfluidity, but that
the superfluid region appears to coincide with the region of equal
density.  The equally exciting work of Partridge \emph{et al.}
\cite{randy}  directly shows phase separation through high
resolution {\em in-situ} images of the atomic clouds.

 In this paper, we study the density profile in the entire BCS-BEC
crossover region within mean field theory.  We use a local density
approximation (LDA) and compare our results with experiments. Our
study shows that the population imbalanced trapped Fermi gas is
generically phase separated into concentric shells. Within our
approximations, each region of space can be in one of several
phases: unpolarized superfluid (S), polarized superfluid (PS),
normal mixture (M), or fully polarized normal (P).  The unpolarized
superfluid coincides with the standard equal-population superfluid
predicted by the BCS-BEC crossover theory
\cite{crossover,model1,model2,model3,cre}. The polarized superfluid,
which in mean-field theory is only found on the BEC side of
resonance, consists of an interpenetrating gas of bosonic molecules
and a fully polarized gas of fermions \cite{dan,pao}. The normal
mixture and fully polarized normal phases both lack superfluidity
and are distinguished by the presence or absence of the minority
species of fermion.  Due to the extremely small portion of the phase
diagram where it is expected to appear, we do not consider the
possibility of an inhomogeneous FFLO phase.  As seen in a number of
recent papers \cite{dan,kinnunen,castorina}, such a phase would
appear on the BCS side of resonance as a subtle structure in the
domain wall separating the S and M regions.

\begin{figure}
\includegraphics[width=\columnwidth]{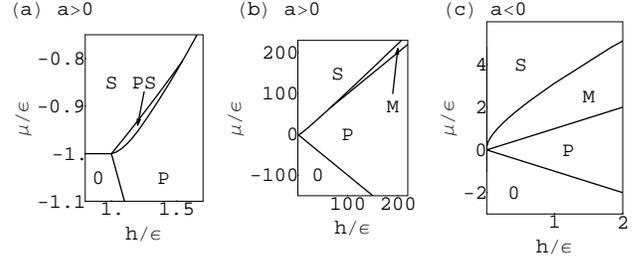}
\caption{Mean-field phase diagram of homogeneous two-component Fermi
gas in the (a), (b) BEC  and  (c)  BCS regimes.  Phases only depend
on the dimensionless ratio of the chemical potentials
$\mu=(\mu_{\uparrow}+\mu_{\downarrow})/2$ and
$h=(\mu_{\uparrow}-\mu_{\downarrow})/2$, and the energy scale of the
interactions,  $\epsilon=\hbar^2/2ma^2$. Panels (a) and (b) differ
only by the scale used.} \label{pd}
\end{figure}

We find three different parameter regimes, distinguished by the
structure of these shells: (1) BCS regime where
$(k_{f}a)^{-1}\lesssim 0$, (2) crossover-BEC regime where $0
\lesssim (k_{f}a)^{-1}\lesssim 1$, and (3) deep-BEC regime where
$(k_{f}a)^{-1}\gtrsim 1$.  The Fermi wave vector is $k_f$ and the
scattering length is $a$.  Depending upon density and polarization,
regime (1) contains two scenarios -- from center to edge one finds
S-M-P or M-P.  In regime (2) one finds S-P at small polarizations
and S-M-P  or M-P for larger. In regime (3) one finds S-PS-P for
small polarizations and PS-P for larger polarizations \cite{note}.
In an appropriately defined thermodynamic limit there is a quantum
phase transition between each of these possibilities as one varies
the parameters of the system.  This behavior should be contrasted
with the smooth crossover physics found in the absence of a
population imbalance.

Our results are consistent with the experiments reported in Ref.
\cite{ketterle}, however, due to the expansion procedure used in
those experiments no quantitative comparison can be made.  We find
partial agreement with the experiments reported in Ref.
\cite{randy}. In particular, for sufficiently large polarizations we
reproduce the values of the axial radius of the superfluid core and
the outer edge of the gas cloud found in the experiments
\cite{randy}. Our total density profiles also closely agree with the
experiments. However, we find that the spatial structure of the
difference between up and down spins differs significantly from
those found in the experiment \cite{randy}. In particular, we show
that despite the large ratio of the trap size to the coherence
length, the experimental data are inconsistent with any LDA that
assumes a harmonic trap, regardless of the equation of state used.

The most significant open theoretical question at the moment is the
possibility of a polarized superfluid region at unitarity (UPS). As
seen from the $\epsilon \rightarrow 0$ behavior of Fig. \ref{pd}(b),
the mean-field calculation predicts that such a region does not
exist. Monte Carlo calculations suggest that such a region may
exist, but are currently not conclusive \cite{carleson}. Based upon
a comparison with our mean-field calculations we argue that the
experimental measure of phase separation (from analyzing the density
profiles) is slightly ambiguous and one may be able to explain the
experiments without recourse to a UPS. On the other hand, the poor
agreement in the radii at small polarizations may support the notion
of a UPS.

Concurrent with the preparation of this manuscript several authors
\cite{kinnunen,new} presented complementary theoretical studies of
the effect of trap potentials on a partially polarized gas. With the
exception of Ref.\cite{dan}, which discusses some qualitative
feature of the trapped gas, prior theoretical work on superfluidity
with mismatched Fermi surfaces has been restricted to either
homogeneous systems \cite{homo,breached,dan,pao,paulo} or trapped
systems in the weakly coupling limit \cite{castorina}.

\begin{figure}
\includegraphics[width=\columnwidth]{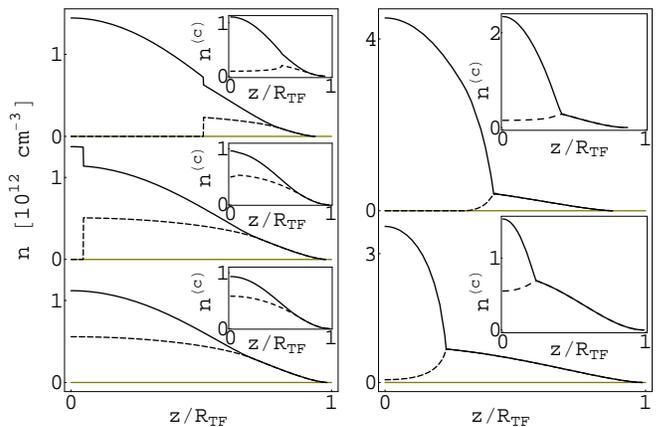}
\caption{ On-axis mean-field density profiles $n(z,\rho=0)$ of zero
temperature harmonically trapped partially polarized Fermi gas,
illustrating all of the cases described in the text.  Solid lines
show total density $n_\uparrow+n_\downarrow$, while dashed lines
show density differences $n_\uparrow-n_\downarrow$.  Left (right)
figures represent the strong coupling BCS (BEC) regime with $k_f
a=-2$, (0.5). The Thomas-Fermi radius  is defined as
$R_{TF}=\sqrt{(2\epsilon_{f})/(m\omega_{z}^{2})}$, where
$\epsilon_{f} = \hbar^{2}k_{f}^{2}/2m =
\hbar\overline{\omega}(6N)^{1/3}$ with average trap frequency
$\overline{\omega}=(\omega_{\perp}^{2}\omega_{z})^{1/3}$.
 Inset shows column density $n^{(c)}=\int d\rho\, n(z,\rho)$ measured in units of [$10^9$ cm$^{-2}$].
 All graphs use $N_{\uparrow}+N_{\downarrow}= 2\times 10^5$ atoms in an axial symmetric trap
 with $\omega_z=2\pi\times 7.2 Hz$, and $\omega_\perp=2\pi\times 350 Hz$.  BCS figures (top to bottom)
 have polarizations $P=(N_{\uparrow}-N_{\downarrow})/(N_{\uparrow}+N_{\downarrow})=0.30, 0.74, 0.80$ and
 BEC figures have $P=0.33, 0.90$.
}\label{profiles}
\end{figure}

\section{Theory}

We restrict ourselves to the wide resonance of $^6$Li  where both of
the available experiments~\cite{ketterle,randy} have been carried
out, and therefore we do not need to explicitly consider the closed
channel of the Feshbach resonance.
 The fermions of different hyperfine states
$\uparrow$ and $\downarrow$ interact through a short-range effective
potential $-u\delta(\vec{r}\prime-\vec{r})$. The system of
$N=N_{\uparrow}+N_{\downarrow}$ atoms is then described by the
Hamiltonian~\cite{model1,model2,model3} $H=H_1+H_2$, with $H_1 =
\sum_{\sigma}\int d^3\vec{r}\,
\psi_{\sigma}^{\dagger}(\vec{r})[-\frac{\hbar^{2}\nabla^2}{2m}-\mu_{0\sigma}+V(\vec{r})]
\psi_{\sigma}(\vec{r})$ containing kinetic and trapping energies and
$H_2=-u\int
d^3\vec{r}\psi^{\dagger}_{\uparrow}(\vec{r})\psi^{\dagger}_{\downarrow}(\vec{r})\psi_{\downarrow}(\vec{r})
\psi_{\uparrow}(\vec{r}) $ containing interactions. The field
operators $\psi_{\sigma}(\vec{r})$ obey the usual fermionic
anticommutation rules, and describe the annihilation of a fermion at
position $\vec{r}$ in the hyperfine state $\sigma$. Parameters $m$,
$\mu_{0\sigma}$ and
$V(\vec{r})=\frac{1}{2}m(\omega_{\perp}^{2}\rho^{2}+\omega_{z}^{2}z^2)$
are the mass, chemical potential, and trapping potential of the
atomic species $\sigma$. We introduce a local chemical potential
$\mu_{\sigma}(\vec{r})=\mu_{0\sigma}-V(\vec{r})$ and treat the
system as locally homogeneous. Without loss of generality, we take
$\uparrow$ to be the majority species of atoms and describe the
system in terms of the spatially independent chemical potential
difference $h=(\mu_{\uparrow}-\mu_{\downarrow})/2 \geq 0$, and the
spatially dependent average chemical potential
$\mu(\vec{r})=(\mu_{\uparrow}+\mu_{\downarrow})/2$. Using the usual
BCS mean-field decoupling, the BCS-Bogoliubov excitation spectrum of
each species is given by $E_{k\sigma}(\vec{r})=\xi_{\sigma}
h+\sqrt{(\epsilon_k-\mu)^2+\Delta^2}$. Here
$\epsilon_k=\hbar^2k^2/2m$ is the kinetic energy,
$\Delta(\vec{r})=u\langle
\psi_{\downarrow}(\vec{r})\psi_{\uparrow}(\vec{r})\rangle$ is the
local superfluid order parameter, and $\xi_{\uparrow}=-1$ and
$\xi_{\downarrow}=+1$.

At zero temperature, the  gap
 $\Delta(\vec{r})$ and the the number densities
$n(\vec{r})=n_{\uparrow}(\vec{r})+n_{\downarrow}(\vec{r})$,
$n_d(\vec{r})=n_{\uparrow}(\vec{r})-n_{\downarrow}(\vec{r})$ satisfy
\begin{eqnarray}\label{reggap}
\frac{-m}{2\pi\hbar^2a} &=&  \int_0^{\infty} \frac{d^3\vec{k}}{(2
\pi)^3}\left(\frac{1}{E_k}-\frac{1}{
\epsilon_k}\right)-\int^{k_{+}}_{k_{-}} \frac{d^3\vec{k}}{(
2\pi)^3}\frac{1}{E_k}\\
n(\vec{r}) &=& \int_0^{\infty} \frac{d^3\vec{k}}{(2
\pi)^3}(1-\frac{\epsilon_k-\mu}{E_k})+\int^{k_{+}}_{k_{-}}(\frac{\epsilon_k-\mu}{E_k})\\
n_d(\vec{r})&=&\frac{1}{(2 \pi)^3}\frac{4 \pi}{3}(k_{+}^3-k_{-}^3).
\end{eqnarray}
We define $E_{k}(\vec{r})=(E_{k\uparrow}+E_{k\downarrow})/2$, and
$k_{\pm}(\vec{r})=(\pm\sqrt{h^2-\Delta^2}+\mu)^{1/2}$. The
ultraviolet divergence associated with the delta-function
interaction has been eliminated~\cite{model1, model2,cre,reg} by
introducing the effective scattering length through
$-{m}/({4\pi\hbar^2a}) = {u}^{-1} - \int_0^{\infty} {d^3\vec{k}}{(2
\pi)^{-3}}/{2\epsilon_k}.$ Notice that $n_d(\vec{r})$ is nonzero as
long as $h>\Delta$ and $\mu>-\sqrt{h^2-\Delta^2}$.

\begin{figure}
\includegraphics[width=\columnwidth]{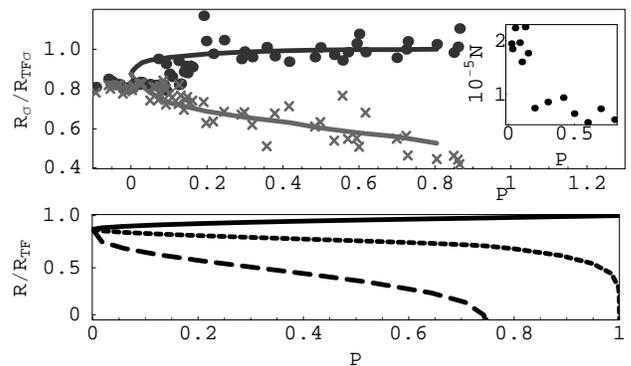}
\caption{ TOP: Radii of the minority and majority components using
experimental parameters from Ref. \cite{randy}. Dots and crosses are
the experimental data for the majority and minority components
respectively. Lines show our theoretical predictions with $k_f
a=20.6$. The inset shows the total number of atoms used for
calculation at each P. The two spin state radii $(R_{\uparrow},
R_{\downarrow})$ are separately scaled by $R_{TF\uparrow}$ and
$R_{TF\downarrow}$, where $R_{TF\sigma}$ is the Thomas-Fermi radius
of a noninteracting one-component cloud with $N_{\sigma}$ atoms.
BOTTOM: Radii of phase boundaries in the BCS regime with $k_f a=-2$.
Solid, short-dashed, and long-dashed lines are the boundaries of
polarized normal, normal mixture, and unpolarized superfluid,
respectively. Input parameters are the same as those of figure
\ref{profiles}, as is the definition of $R_{TF}$. }\label{radii}
\end{figure}

At each point in space one finds $\Delta$ by solving Eq.
(\ref{reggap}) at fixed $\mu$ and $h$.  As a nonlinear equation,
there are multiple solutions -- some of which correspond to energy
minima, some to saddle points. For example, the Sarma state
\cite{breached, paulo, sarma} appears as a saddle point. We take the
lowest energy solution, producing the phase diagram illustrated in
Fig.~\ref{pd}.  Within the cloud $h$ is uniform, while $\mu$ varies
monotonically from the center to edge of the cloud. The
configurations of local phases, as described in the Introduction,
are found by following vertical lines in the figure. One determines
$\mu_0$ and $h$ by imposing a constraint on the total number of
particles $N = \int d^3\vec{r}\, n(\vec{r})$ and the polarization
$P=(N_{\uparrow}-N_{\downarrow})/N$.  Typical density profiles are
shown in Fig.~\ref{profiles}. Evolution of the radii of phase
boundaries are shown in Fig.~\ref{radii}.

\section{Experiments}

\begin{figure}
\includegraphics[width=\columnwidth]{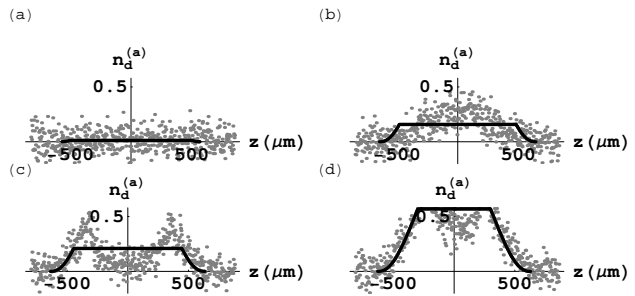}
\caption{ Doubly integrated axial density difference
$n_d^{(a)}(z)=2\pi\,\int d\rho\,\rho\,
[n_{\uparrow}(z,\rho)-n_{\downarrow}(z,\rho)]$ of zero temperature
harmonically trapped partially polarized Fermi gas in units of
$[10^6 cm^{-1}]$. Figures (a), (b), (c), and (d) represent
polarization $P= 0.01$, 0.09, 0.14, and 0.53 respectively. All
graphs use the experimental parameters from Ref. \cite{randy} and
the gray points are the experimental data.}\label{axial}
\end{figure}

\begin{figure}
\includegraphics[width=\columnwidth]{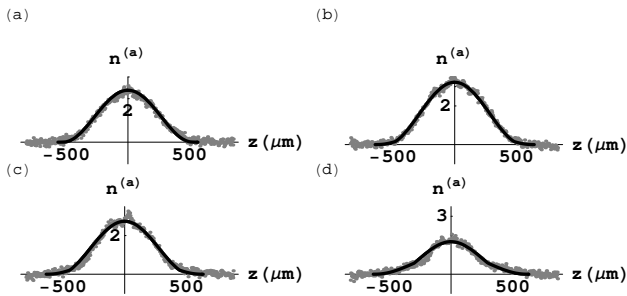}
\caption{ Comparison of the axial density $n^{(a)}(z)=2\pi\,\int
d\rho\,\rho\, [n_{\uparrow}(z,\rho)+n_{\downarrow}(z,\rho)]$ in
units of $[10^6 cm^{-1}]$ with experimental data of Ref.
\cite{randy}. The labels are the same as those of Fig.
\ref{axial}.}\label{td}
\end{figure}

There are profound differences in the full three-dimensional density
profile of the experimental cloud of Ref. \cite{randy} and our
predictions. These differences are best seen by looking at the {\em
axial} density profile $n_d^{(a)}(z)=2\pi\,\int d\rho\,\rho\,
n_d(z,\rho)$, found by integrating the densities over both
transverse directions. As shown in Fig.~\ref{axial}, we find a
monotonic axial profile, while at polarizations above $P\approx0.1$,
Partridge {\em et al.} experimentally see a non monotonic density
difference, with a dip in the center and horns on the edges. Despite
these horns, our calculation of the total axial density
$n^{(a)}(z)=2\pi\,\int d\rho\,\rho\, n(z,\rho)$ agrees extremely
well with the experimental data (Fig. \ref{td}).

Assuming LDA in a harmonic trap, the density is only a function of
$\mu(r)=\mu_{0}
-m\omega_{\perp}^{2}\rho^{2}/2-m\omega_{z}^{2}z^{2}/2$. [As before,
$\mu_{0}=(\mu_{0\uparrow}+\mu_{0\downarrow})/2$.] One can therefore
write $n_d^{({a})}(z)= \frac{2\pi}{m\omega_{\perp}^{2}}f(\mu_{0} -
m\omega_{z}^{2}z^{2}/2),$ where $f(\bar\mu) =
\int_{-\infty}^{\bar\mu}d\mu n_{d}(\mu).$ Since $n_d>0$ one has
$f(\bar \mu)$ is monotonic and $n_d^{({a})}(z)$ must decrease
monotonically as $z$ increases from 0. In other words the horns seen
in the experimental density differences are not consistent with the
LDA. This result is based solely on the local density approximation
and harmonic trapping -- it does not require mean-field theory. We
caution that this theorem only applies to the doubly integrated
axial density, and not to the column density.

Even if mean-field theory breaks down, these experiments should be
well described by a local density approximation.  The only relevant
microscopic scale near unitary is the Fermi temperature $T_f\approx
400nK$, which is over 20 times larger than the quantization scale of
the harmonic trap $\hbar \omega_\perp/k_B\approx 17 nK$, which is a
characteristic scale for density variations. We are currently
investigating the possibility that surface tension in the domain
wall between the polarized and superfluid regions may be distorting
the shape of the boundary, leading to the observed density profiles.

Further experimental and theoretical work is needed to settle this
issue. It is particularly important because Partridge {\em et al.}
interpret the appearance of horns as a transition between a unitary
polarized superfluid and a phase separation between superfluid and
normal regions.  Since the LDA predicts that the horns do not exist
in a phase-separated cloud, we caution strongly against taking their
disappearance as evidence for a polarized superfluid.

\section{ Acknowledgments}

This work was supported by NSF grant No. PHY-0456261, and by the
Alfred P. Sloan Foundation. We are grateful to R. Hulet and W. Li
for very enlightening discussions, and for sending us the data in
Figs. \ref{radii}, \ref{axial}, and \ref{td}. We thank D. E. Sheehy
and L. Radzihovsky for the critical comments on the manuscript.


\begin{references}

\bibitem{fflo} P. Fulde and R. A. Ferrell, Phys. Rev. \textbf{135}, A550
(1964) and A.I. Larkin and Yu.N. Ovchinnikov, Zh. Eksp. Teor. Fiz
47, 1136 (1964) [Sov. Phys. JETP 20, 762 (1965)].

\bibitem{nm}A. Sedrakian and U. Lombardo, Phys. Rev. Lett. \textbf{84}, 602
(2000); J. A. Bowers and K. Rajagopal, Phys. Rev. D \textbf{66},
065002 (2002); I. Shovkovy and M. Huang, Phys. Lett. B \textbf{564},
205 (2003); R. Casalbuoni and G. Nardulli, Rev. Mod. Phys.
\textbf{76}, 263 (2004).

\bibitem{feshbach} H. Feshbach, Ann.Phys. \textbf{19}, 287 (1962).

\bibitem{ketterle} Martin W. Zwierlein, André Schirotzek, Christian H. Schunck, Wolfgang
Ketterle, Science, \textbf{311}, 492 (2006).


\bibitem{randy} Guthrie B. Partridge, Wenhui Li, Ramsey I. Kamar, Yean-an Liao, and
Randall G. Hulet, Science, \textbf{311}, 503 (2006).


\bibitem{crossover}A. J. Leggett, in Modern Trends in the Theory of Condensed Matter, edited by
A. Pekalski and R. Przystawa (Springer-Verlag, Berlin, 1980); P.
Nozieres and S. Schmitt-Rink, J. Low Temp. Phys. \textbf{59}, 195
(1985); D. M. Eagles, Phys. Rev. \textbf{186}, 456 (1969); A.
Tokumitsu, K. Miyake and, K. Yamada, Phys. B \textbf{47}, 11988
(1993); J. R. Engelbrecht, M. Randeria and C. A. R. Sa de Melo,
Phys. Rev. B \textbf{55}, 15153 (1997).

\bibitem{model1} Y. Ohashi and A. Griffin, Phys. Rev. Lett. \textbf{89}, 130402 (2002).

\bibitem{model2}J. N. Milstein, S. J. J. M. F. Kokkelmans, and M. J. Holland,
Phys. Rev. A \textbf{66}, 043604 (2002).

\bibitem{model3} E. Timmermans, K. Furuya, P. W. Milonni, and A. K. Kerman, Phys.
Lett. A \textbf{285}, 228 (2001); M. Holland, S. J. M. F.
Kokkelmans, M. L. Chiofalo, and R. Walser, Phys. Rev. Lett.
\textbf{87}, 120406 (2001); A. Perali, P. Pieri, and, G. C.
Strinati, Phys. Rev. A \textbf{68}, 031601(R) (2003); T. Kostyrko
and J. Ranninger, Phys. Rev. B \textbf{54}, 13105 (1996).

\bibitem{cre} C.A.R. Sá de Melo, M. Randeria, and J.R. Engelbrecht, Phys. Rev. Lett. \textbf{71}, 3202 (1993);
Y. Ohashi1 and A. Griffin, Phys. Rev. A \textbf{67}, 033603 (2003);
Y. Ohashi1 and A. Griffin, Phys. Rev. A \textbf{67}, 063612 (2003).

\bibitem{dan} D.E. Sheehy and L. Radzihovsky, PRL \textbf{96}, 060401
(2006).


\bibitem{pao} C.-H. Pao, Shin-TzaWu, and S.-K. Yip, preprint, cond-mat/0506437.

\bibitem{kinnunen} J. Kinnunen, L. M. Jensen, and P. Torma, Phys. Rev. Lett. 96, 110403
(2006).

\bibitem{castorina} P. Castorina,
M. Grasso, M. Oertel, M. Urban, and D. Zappal`a, Phys. Rev. A
\textbf{72}, 025601 (2005); T. Mizushima, K. Machida, and M.
Ichioka, Phys. Rev. Lett. \textbf{94}, 060404 (2005).


\bibitem{note} This PS-P shell structure was discussed in reference
\cite{dan}.

\bibitem{carleson} J. Carlson and S. Reddy, Phys. Rev. Lett. \textbf{95}, 060401 (2005).

\bibitem{new} P. Pieri, and G.C. Strinati, PRL 96, 150404 (2006); W. Yi, and L. -M. Duan, Phys. Rev. A 73, 031604(R) (2006); F.
Chevy, Phys. Rev. Lett. 96, 130401
 (2006); H. Caldas, preprint,
cond-mat/0601148; M. Haque and H.T.C. Stoof, preprint,
cond-mat/0601321; Tin-Lun Ho and Hui Zhai, preprint
cond-mat/0602568; Zheng-Cheng Gu, Geoff Warner and Fei Zhou,
preprint cond-mat/0603091 ;  M. Iskin and C. A. R. Sa de Melo,
preprint cond-mat/0604184;  T. Paananen, J.-P. Martikainen, P.
Torma, preprint cond-mat/0603498.

\bibitem{homo} R. Combescot, Europhys. Lett. \textbf{55}, 150 (2001); H.
Caldas, Phys. Rev. A \textbf{69}, 063602 (2004); A. Sedrakian, J.
Mur-Petit, A. Polls; H. M¨uther, Phys. Rev. A \textbf{72}, 013613
(2005); U. Lombardo, P. Nozi`eres, P. Schuck, H.-J. Schulze, and A.
Sedrakian, Phys. Rev. C \textbf{64}, 064314 (2001);  D.T. Son and
M.A. Stephanov, cond-mat/0507586; L. He, M. Jin and P. Zhuang,
cond-mat/0601147.

\bibitem{breached}W.V. Liu and F. Wilczek, Phys. Rev. Lett. \textbf{90}, 047002
(2003).

\bibitem{paulo} P. F. Bedaque, H. Caldas, and G. Rupak, Phys. Rev. Lett. \textbf{91}, 247002
(2003).

\bibitem{sarma}G. Sarma, J. Phys. Chem. Solids \textbf{24}, 1029
(1963).


\bibitem{reg} S.J.J.M.F. Kokkelmans, J.N. Milstein, M.L. Chiofalo, R. Walser,
and M.J. Holland, Phys. Rev. A \textbf{65}, 053617 (2002).


\end{references}
\end{document}